\documentclass[seceq,supplement]{ptptex}

\usepackage{graphicx}

\title{
Heavy-heavy-light quark potential in two approaches
}

\author{
Arata \textsc{Yamamoto}$^{1}$, Hideo \textsc{Suganuma}$^{1}$ and Hideaki \textsc{Iida}$^{2}$%
}

\inst{
$^{1}$Department of Physics, Faculty of Science, Kyoto University,\\Kitashirakawa, Sakyo, Kyoto 606-8502, Japan\\
$^{2}$Yukawa Institute for Theoretical Physics, Kyoto University,\\Kitashirakawa, Sakyo, Kyoto 606-8502, Japan
}

\abst{
We perform the first study about the heavy-heavy-light quark potential in lattice QCD and a potential model.
We find that the inter-two-quark confining force is reduced by valence quark motional effects compared to the string tension.
}

\begin{document}

\maketitle

\section{Introduction}
There are a large number of studies about the interquark potential in hadrons.
The interquark potential is mainly generated by gluon dynamics reflecting the SU(3) gauge symmetry.
In particular, as many lattice QCD studies, the static and quenched approximation produces the only gluonic potential.
However, realistic hadrons include not only gluons but also quarks, and quarks affect the interquark potential.
We can consider two types of the quark effects, dynamical quark effects and valence quark effects.
An example of the dynamical quark effects is the string breaking, which is obtained by unquenched calculation.
In this study, we focus on the motional effects of finite-mass valence quarks, that is, valence quarks are not static but have a finite mass.

To investigate such quark effects, we define the heavy-heavy-light ($QQq$) quark potential $V_{QQq}(R)$.
It is defined as the energy of $QQq$ systems in terms of the inter-heavy-quark distance $R$.
The $QQq$ system is constructed from two heavy quarks which are infinitely heavy and spatially fixed, and one light quark which is finite mass and moving around.
Since this $QQq$ potential effectively includes the light-quark motional effect, we can investigate the finite-mass valence quark effects to the interquark force.
This $QQq$ system is also related to the doubly charmed baryon, which has been recently discovered at the SELEX experiment\cite{Ma02,Oc05}.

To calculate the $QQq$ potential, we adopt the two approaches, SU(3) lattice QCD\cite{YaUP} and a non-relativistic potential model\cite{Ya08}.
Lattice QCD is the first principle calculation based on QCD.
The potential model, or the quark model, is one of the most successful analytical approaches for the low-energy hadron physics.

\begin{figure}[t]
\begin{minipage}{0.49\textwidth}
\begin{center}
\includegraphics[scale=0.5]{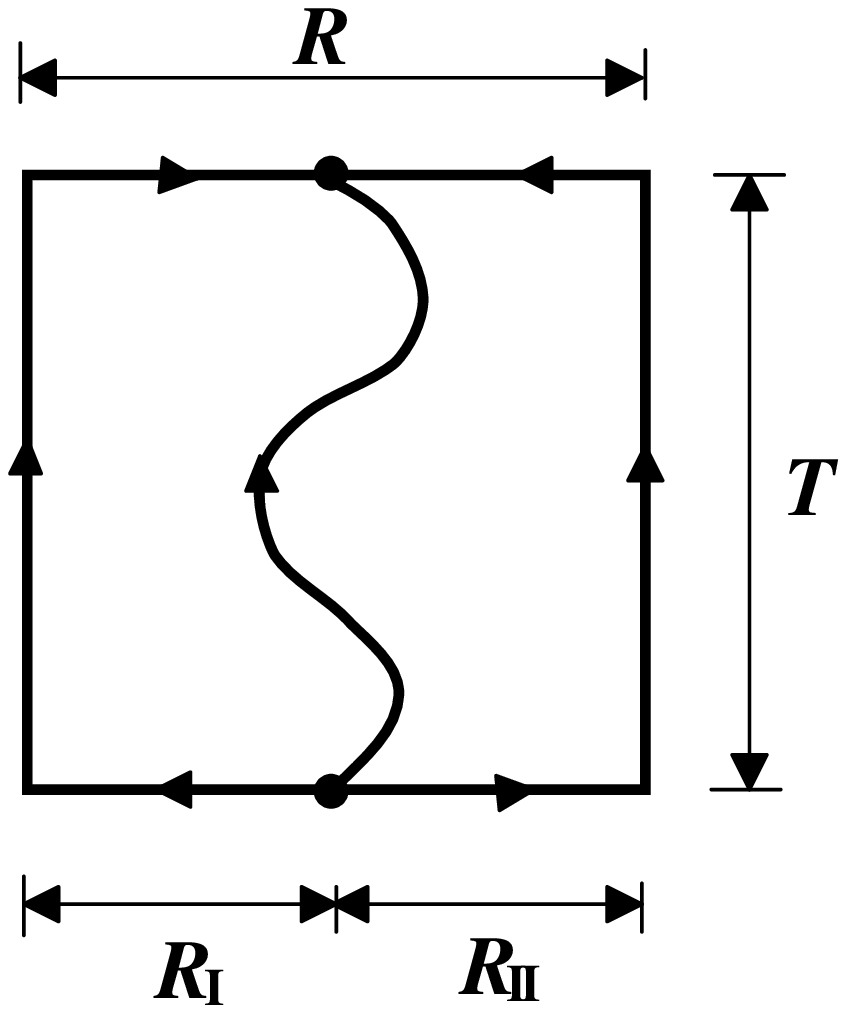}
\caption{\label{fig1}
The $QQq$ Wilson loop.
The wavy line represents the light-quark propagator and the straight line the heavy-quark trajectory.
}
\end{center}
\end{minipage}
\begin{minipage}{0.49\textwidth}
\begin{center}
\includegraphics[scale=0.9]{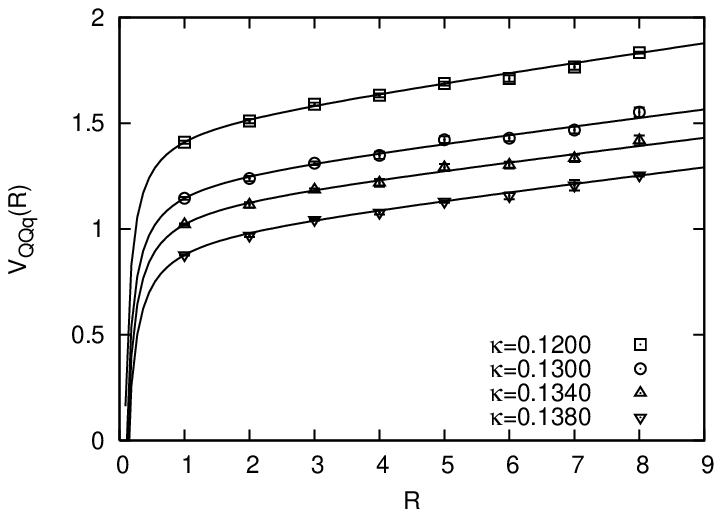}
\caption{\label{fig2}
The Coulomb gauge fixed $QQq$ potential $V_{QQq}$ in the lattice QCD approach.
All the scales are measured in the lattice unit $a\simeq 0.10$ fm.
}
\end{center}
\end{minipage}
\end{figure}

\section{Lattice QCD approach}
In lattice QCD, the $QQq$ potential is extracted from the expectation value of the $QQq$ Wilson loop $W_{QQq}(R,T)$.
As shown in Fig.~\ref{fig1}, the $QQq$ Wilson loop is constructed from the heavy-quark trajectories and the light-quark propagator.
The calculation is performed on $16^4$ quenched lattice with $\beta=6.0$.
For the statistical error reduction, we apply the wall-to-wall propagator for the light-quark propagator, using the Coulomb gauge\cite{YaUP}.
Since the Coulomb gauge fixing can give a gauge variant contribution to the short-range physics, we investigate the long-range behavior of the potential, which is empirically unchanged by the Coulomb gauge fixing\cite{Gr03}.
Such a procedure surprisingly suppresses the statistical error.
 
As the analogy of the $Q\bar Q$ potential, the $QQq$ potential is fitted with
\begin{eqnarray}
V_{QQq}(R)=\sigma _{\rm eff}R-\frac{A_{\rm eff}}{R}+C_{\rm eff}.
\label{VQQq}
\end{eqnarray}
As shown in Fig.~\ref{fig3} and Table \ref{tab1}, this function is surprisingly suitable for the $QQq$ potential.
Compared to the string tension $\sigma$ of the static $Q\bar Q$ or $3Q$ potential, the effective string tension $\sigma _{\rm eff}$ is about 10-20\% reduced,
\begin{eqnarray}
\sigma_{\rm eff} < \sigma \simeq  0.89\ {\rm GeV/fm},
\end{eqnarray}
except for $\kappa =0.1200$ ($M_q=1.5$ GeV).
This result means that the inter-heavy-quark confining force is reduced due to the light-quark existence.

As for the light-quark mass dependence, the effective string tension increases when the light-quark mass increases.
In the infinite light-quark mass limit, the $QQq$ system becomes the static $3Q$ system, and then $\sigma _{\rm eff}$ would approach $\sigma$.

\begin{table}[b]
\begin{minipage}{0.6\textwidth}
\begin{center}
\caption{\label{tab1}
The lattice QCD result with the light-quark hopping parameter $\kappa$ of the $O(a)$-improved Wilson fermion\cite{El97}.
The constituent quark mass $M_q$ is estimated as the half of the $\rho$ meson mass.
}
\begin{tabular}{ccccc}
\hline\hline
$\kappa$ & $M_q$ [GeV] & $\sigma_{\rm eff}$ [GeV/fm] & $A_{\rm eff}$ & $\chi ^2/N_{\rm dof}$\\
\hline
0.1380 &0.5 & 0.73(3) & 0.13(1) & 1.16\\
0.1340 &0.7 & 0.73(8) & 0.13(2) & 1.11\\
0.1300 &1.0 & 0.75(8) & 0.13(2) & 1.18\\
0.1200 &1.5 & 0.89(4) & 0.12(2) & 1.31\\
\hline\hline
\end{tabular}
\end{center}
\end{minipage}
\hspace{5pt}
\begin{minipage}{0.35\textwidth}
\begin{center}
\caption{\label{tab2}
The potential model result.
The error from the finite volume and discretization is estimated within a few percent.
}
\begin{tabular}{ccc}
\hline\hline
$M_q$ [GeV]& $\sigma_{\rm eff}$ [GeV/fm] & $A_{\rm eff}$\\
\hline
0.33& 0.73 & 0.12\\
0.50& 0.75 & 0.12\\
1.0 & 0.80 & 0.12\\
2.0 & 0.84 & 0.12\\
\hline\hline
\end{tabular}
\end{center}
\end{minipage}
\end{table}

\section{Potential model approach}

In the potential model, the $QQq$ system is treated as a non-relativistic constituent quark system.
The Hamiltonian of this system is written as
\begin{eqnarray}
H=M_q-\frac{1}{2M_q}\frac{\partial^2}{\partial \vec{r}_3^2}+V(\vec{r}_1,\vec{r}_2,\vec{r}_3),
\end{eqnarray}
where the subscripts 1, 2, and 3 mean the two heavy quarks and the light quark, respectively, and $M_q$ is the light-quark constituent mass.
For the three-quark interaction, we adopt not the simple sum of the two-body force in ordinary quark models, but the three-body force based on the static $3Q$ potential in quenched lattice QCD\cite{Ta0102} as
\begin{eqnarray}
V(\vec{r}_1,\vec{r}_2,\vec{r}_3)=\sigma L_{\rm min}-\sum _{i< j}\frac{A_{3Q}}{|\vec{r}_i-\vec{r}_j|},
\end{eqnarray}
where $\sigma = 0.89$ \rm GeV/fm and $A_{3Q} = 0.13$.
The color flux-tube length $L_{\rm min}$ is given as the length minimally connecting the three quarks.
Once the heavy-quark coordinates are fixed, we can calculate the energy $E(R)$ of the system only through the light-quark wave function.
By minimizing $E(R)$ in discretized space, we determine the ground-state light-quark wave function $\psi (\vec{r}_3)$ and the ground-state $QQq$ potential $V_{QQq}(R)$.
For the energy variational calculation, we adopt the ``renormalization group inspired variational calculation"\cite{Ya08}.

\begin{figure}[t]
\begin{minipage}{0.49\textwidth}
\includegraphics[scale=1.4]{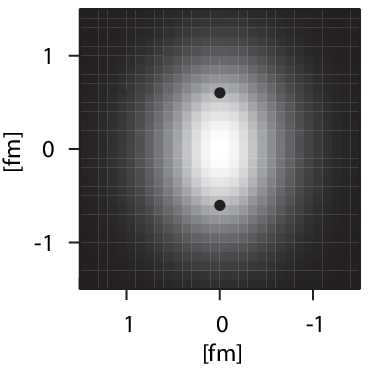}
\caption{\label{fig3}
The light-quark spatial distribution $|\psi (\vec{r}_3)|^2$ with $M_q=330$ MeV.
Black circles denote the heavy-quark positions.
}
\end{minipage}
\begin{minipage}{0.49\textwidth}
\includegraphics[scale=0.9]{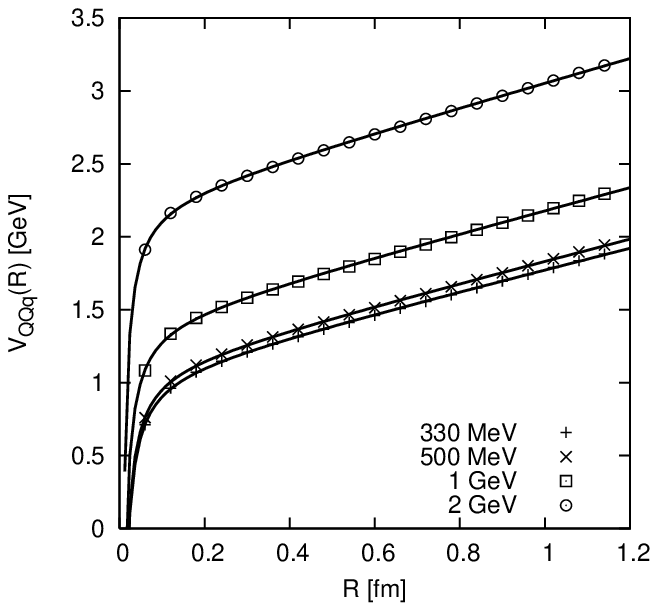}
\caption{\label{fig4}
The $QQq$ potential $V_{QQq}(R)$ in the potential model approach.
The solid curves are the best-fit functions of Eq.~(\ref{VQQq}).
}
\end{minipage}
\end{figure}

An example of the light-quark spatial distribution is shown in Fig.~\ref{fig3}.
The $QQq$ potential is fitted with Eq.~(\ref{VQQq}) in the range of $R\le 1.2$ fm, and the result is shown in Fig.~\ref{fig4} and Table \ref{tab2}.
We can see the same behavior in lattice QCD, i.e., the reduction of the effective string tension and its $M_q$-dependence.
To compare with the lattice QCD result, we calculate in the potential model with the same condition, $M_q=1$ GeV and $R\le 0.8$ fm, and find $\sigma _{\rm eff} \simeq 0.76$ GeV/fm.
Thus, the calculation of the simple potential model almost reproduces the result of lattice QCD.

In this potential model, we can calculate the expectation value $\langle L_{\rm min} \rangle$ of the flux-tube length from the obtained light-quark wave function.
We find that $\langle L_{\rm min} \rangle$ is almost a linear function of the inter-heavy-quark distance $R$, i.e., $\langle L_{\rm min} \rangle \simeq b_0+b_1R$, and that the essential point is the relation between the flux-tube length and the inter-heavy-quark distance, as
\begin{eqnarray}
\langle \sigma L_{\rm min} \rangle \simeq (b_1\sigma )R+b_0\sigma \simeq \sigma_{\rm eff}R +{\rm const.}
\label{S3QLmin}
\end{eqnarray}
Thus, $\sigma_{\rm eff}$ is reduced as $\sigma_{\rm eff}\simeq b_1\sigma$, and $b_1$ means the reduction rate of $\sigma_{\rm eff}$.

Note that the relation between the flux-tube length and the inter-heavy-quark distance can be generally more complicated, reflecting the light-quark dynamics.
We confirm that, in the region of $R\le 2.4$ fm, the $QQq$ potential deviates from Eq.~(\ref{VQQq}) and $\sigma _{\rm eff}$ is an increasing function of $R$.
In the $R \to \infty$ limit, $R$ approaches $\langle L_{\rm min}\rangle$, so that $\sigma _{\rm eff}$ would approach $\sigma$.

\section{Conclusion}

From the two approaches, we can conclude that the effective string tension between the two heavy quarks is reduced by the light-quark effect, compared to the string tension. 
The reduction of the effective string tension originates from the geometrical relation between the inter-heavy-quark distance $R$ and the flux-tube length $L_{\rm min}$.
This relation is determined by the light-quark dynamics, and reflects the characteristic flux-tube structure in baryons\cite{Ta0102,Ic03}.
Such a light-quark effect is expected to hold not only for $QQq$ systems but also for ordinary baryons.
In addition, also in multi-quark systems, the inter-two-quark confining force would be affected by the existence of other light quarks.

\section*{Acknowledgements}
H.~S.~was supported by a Grant for Scientific Research [(C) No.19540287] in Japan.
The lattice QCD calculations are done on NEC SX-8R at Osaka University.

\end{document}